\documentclass[twocolumn,showpacs,preprintnumbers,amsmath,amssymb]{revtex4}

\usepackage{graphics}
\usepackage{graphicx}
\usepackage{dcolumn}
\usepackage{bm}
\usepackage{amssymb}
\begin{document}
\preprint{APS/123-QED}

\title{Complex microwave conductivity of Pr$_{1.85}$Ce$_{0.15}$CuO$_{4-\delta}$ thin films using a cavity perturbation
method}
\author{Guillaume C\^ot\'e}
\author{Mario Poirier}
\author{Patrick Fournier}
\affiliation{Regroupement Qu\'eb\'ecois sur les Mat\'eriaux de
Pointe, D\'epartement de Physique, Universit\'e de Sherbrooke,
Sherbrooke, Qu\'ebec,Canada J1K 2R1}

\date{\today}

\begin{abstract}
We report a study of the microwave conductivity of electron-doped
Pr$_{1.85}$Ce$_{0.15}$CuO$_{4-\delta}$ superconducting thin films
using a cavity perturbation technique. The relative frequency
shifts obtained for the samples placed at a maximum electric
field location in the cavity are treated using the high
conductivity limit presented recently by Peligrad $\textit{et}$
$\textit{al.}$ Using two resonance modes, TE$_{102}$ (16.5 GHz)
and TE$_{101}$ (13 GHz) of the same cavity, only one adjustable
parameter $\Gamma$ is needed to link the frequency shifts of an
empty cavity to the ones of a cavity loaded with a perfect
conductor. Moreover, by studying different sample configurations,
we can relate the substrate effects on the frequency shifts to a
scaling factor. These procedures allow us to extract the
temperature dependence of the complex penetration depth and the
complex microwave conductivity of two films with different
quality. Our data confirm that all the physical properties of the
superconducting state are consistent with an order parameter with
lines of nodes. Moreover, we demonstrate the high sensitivity of
these properties on the quality of the films.

\end{abstract}

\maketitle
\section{INTRODUCTION}
In high-$T_c$ cuprate superconductors the study of the pairing
symmetry remains a very active field of research and microwave
measurements using resonant structures have shown to be
particularly useful to detect its signatures. Indeed, when a
microwave resonant cavity is loaded with a conducting sample, we
can measure shifts of the resonance frequency and of the quality
factor Q, which can be related to the real and imaginary parts of
the complex conductivity $\Tilde{\sigma} = \sigma_1 - i \sigma_2$
\cite{CASIMIR,WALDRON,HALBRITTER}. Based on the two-fluid model
for a superconductor, condensed pairs and quasiparticles both
contribute to $\Tilde{\sigma}$ which in turn determine the
temperature dependence of the magnetic penetration depth $\lambda
(T)$ and the quasiparticle relaxation rate
$T_1^{-1}(T)$\cite{BONN1993}.

The $d$-wave symmetry is now well established in hole-doped
cuprates superconductors
\cite{HARDY1993,JACOBS1995,LEE1996,KITANO1998}. However, the
situation remains controversial in electron-doped ones
$R_{2-x}Ce_xCuO_{4-\delta}$ $(R = Pr, Nd, Sm)$
\cite{KOKALES,PROZOROV,SNEZHKO}. When single crystals are used,
residual microwave absorption is generally measured in the
superconducting state and this could impede any precise
determination of $\lambda (T)$. This residual absorption is likely
due to an inhomogeneous oxygen reduction process over the volume
of the crystal, a reduction which is mandatory to induce the
superconducting state. In this respect, thin films of these
materials appear more adapted to microwave experiments. Indeed,
high-quality $R_{2-x}Ce_xCuO_{4-\delta}$ thin films can be grown
by pulsed laser ablation on appropriate substrates. Their small
thickness and an in-situ post-annealing ensures a homogeneous
reduction over the complete volume of the sample. There is,
however, an important difference between microwave measurements
of thick samples (single crystals) and thin samples (thin films)
which has not been completely overcome yet experimentally.

The perturbation of a microwave cavity by a small sample of
variable conducting properties is an old problem that has been
treated with different approaches for thin superconducting films
\cite{SCHAUMBURG1992,SCHAUMBURG1994,PELIGRAD1998}. However, a
general procedure proposing to treat the cavity loaded with a
perfect conductor sample as the unperturbed state and to find the
shifts when the sample becomes a non-perfect conductor, appears
particularly promising for high conductivity thin films. A
general expression has been obtained for two intracavity
arrangements and analytical solutions were found for the slab
geometry of a thin film \cite{PELIGRAD1998,PELIGRAD2001}. The
general solution for the complex frequency shift reduces to the
Shchegolev formula \cite{SHCHEGOLEV} in the depolarization limit
when the microwave electric field fully penetrates the sample,
and to the skin depth regime when the microwave penetration depth
becomes smaller than the size of the sample and the complex
frequency shift is related to the surface impedance \cite{KLEIN}.
Thin films having different thicknesses and conductivity can thus
be treated anywhere between the depolarization regime and the
regime close to the skin depth one \cite{PELIGRAD2001}. However,
the high-temperature superconducting thin films are grown on
dielectric substrates having relative permittivity $\epsilon_r$
around 20. Hence, when the sample is introduced in the microwave
cavity, the substrate changes the field configuration
asymmetrically around the sample and modifies the frequency
shifts. A method has been proposed to mimic this asymmetric
solution by introducing the concept of a fictitious substitute
sample. This method has been tested on thin films of hole-doped
cuprates mounted in a cylindrical cavity resonating in the
TE$_{111}$ mode \cite{PELIGRAD2001}.

In this paper we present microwave measurements on optimally doped
Pr$_{1.85}$Ce$_{0.15}$CuO$_{4-\delta}$ (PCCO) thin films grown on
a $LaAlO_3$ substrate mounted in the electric field of a
rectangular cavity. We use the general solution for the complex
frequency shift in microwave electric field in the high
conductivity limit \cite{PELIGRAD1998}. However, to deduce the
complex conductivity from the measured frequency shifts, the
fictitious substitute sample method could not be applied to our
measurements to take into account the impact of the dielectric
substrate. We rather propose a modified version of the method
presented by Peligrad \textit{et al.} exploiting the measurements
of the same sample at two different resonance frequencies and by
analyzing different substrate configurations. Then, the frequency
shifts are transformed into complex penetration depth and
microwave conductivity data. These results obtained on two films
of different quality are discussed in terms of a two-fluid model.
The characteristics of the superconducting state are found to be
very similar to the ones obtained in the hole-doped cuprates.

\section{EXPERIMENT}
The PCCO thin films were grown by pulsed laser deposition on
standard $LaAlO_3$ substrates and their thickness was measured
with a scanning electronic microscope at grazing incidence giving
a precision of 5 nm. Prior to the microwave measurements, the
films were also characterized by four-probe resistivity and AC
magnetic susceptibility measurements were performed in a PPMS from
Quantum Design between 4 and 300 K. For the two films
investigated here, we give in Table 1 the values of the thickness
$\it d$, the critical temperature $T_c$ and the minimum value of
the resistivity $\rho_{min}$ appearing between 14 and 15 K when a
magnetic field of 9T suppresses the superconducting state
\cite{FOURNIER1998}. $T_c$ is determined from the microwave
experiment as indicated in the next section.

The microwave measurements were performed with a cavity
perturbation technique in the transmission mode. We used a
rectangular copper cavity for which the TE$_{102}$ (16.5 GHz) and
TE$_{101}$ (13 GHz) resonance modes could be excited. A
synthesized sweeper and a scalar network analyzer are used to
acquire the resonance curve of each mode after averaging and
smoothing. A lorentzian fit of the curve then yields the
resonance frequency $f_0$ and the half width at half maximum
$(1/2Q)$. This is done for the cavity with and without the sample
installed in the maximum electric field position in order to
obtain the relative frequency shifts $\Delta f / f$ and
$\Delta(1/2Q)$. A micrometer screw allows for the displacement of
the sample in and out of the cavity at low temperatures. With this
set-up the noise level is usually better than 1 kHz. The sample
(thin film + substrate) has a slab geometry with typical
dimensions 2 x 0.2 x 0.5 mm$^3$, 0.5 mm being the thickness of
the substrate. The sample and cavity are inserted in a variable
temperature insert that permits to scan the temperature between 2
and 300 K with a Cernox sensor and a Lake Shore temperature
controller.

\begin{table}[h]
\caption{\label{table1}Physical parameters of two PCCO films.}
\begin{ruledtabular}
\begin{tabular}{cccc}
$Film$ & $\it d (10^{-9} m)$ & $T_c (K)$ & $\rho_{min} (\mu \Omega\ cm)$ \\
\hline
 &  &  &  \\
A & 225(5) & 22.5(1)  & 101 $\pm$ 6 \\
&  &  &  \\
B & 170(5) & 21.5(1) & 310 $\pm$ 40 \\
\end{tabular}
\end{ruledtabular}
\end{table}

\section{RESULTS AND DISCUSSION}
The relative frequency shifts presented in this section will be
treated with the formula valid in the high conductivity limit for
a slab oriented with its length along the maximum electric field
\cite{PELIGRAD1998}. Because our frequency shifts are measured
relative to an empty cavity when the formula implies shifts
relative to a cavity perturbed by a perfect conductor
(unperturbed state), we first describe the appropriate correction
to transform one into the other. Then, we explain how the
substrate effects are included in the treatment. Finally, the
temperature dependence of the complex in-plane penetration depth
$\Tilde{\lambda}$ = $\lambda_1$ - i$\lambda_2$ obtained from the
frequency shifts and the subsequent complex conductivity data
$\Tilde{\sigma} = 1 / i \mu_0 \omega {\Tilde{\lambda}}^2$ for two
PCCO films are presented and discussed.

\subsection{The Unperturbed State}

Theoretically, when a slab of a perfect conductor is introduced
in the electric field of the cavity, the relative shift in
resonance frequency $\Delta f/f$ depends only on a geometrical
factor $\Gamma$ = $\alpha/N$, where $\alpha$ and $N$ are
respectively the filling and the depolarization factors. Then,
the difference in shifts between the empty cavity and the
unperturbed state should be this constant $\Gamma$. Since there
is no dissipation in a perfect conductor, there is no change
induced in the $\Delta(1/2Q)$ data. An example of these shifts
for sample A, corrected for this $\Gamma$ factor, is presented in
Fig. 1. The shift $\Delta(1/2Q)$ is almost zero in the
superconducting state but increases very rapidly when $T_c$ is
approached; a monotonous increase is further observed in the
normal state. The shift $\Delta f/f$ is negative in the
superconducting state with a dip appearing just below $T_c$ which
is defined as the temperature where the rate of increase is
maximum (indicated by an arrow); in the normal state above $T_c$,
the small shift presents a flat temperature dependence for the 13
GHz mode while a monotonous increase is observed at 16.5 GHz. The
temperature and frequency dependence of these relative shifts is
fully consistent with a superconducting sample in the high
conductivity limit \cite{PELIGRAD1998}, except for $\Delta f/f$
above $T_c$ for the 16.5 GHz mode. The absence of a flat
temperature dependence can be explained by a temperature
dependent $\Gamma(T)$ for this particular mode.

\begin{figure}[H,h]
\includegraphics[width=8.5cm]{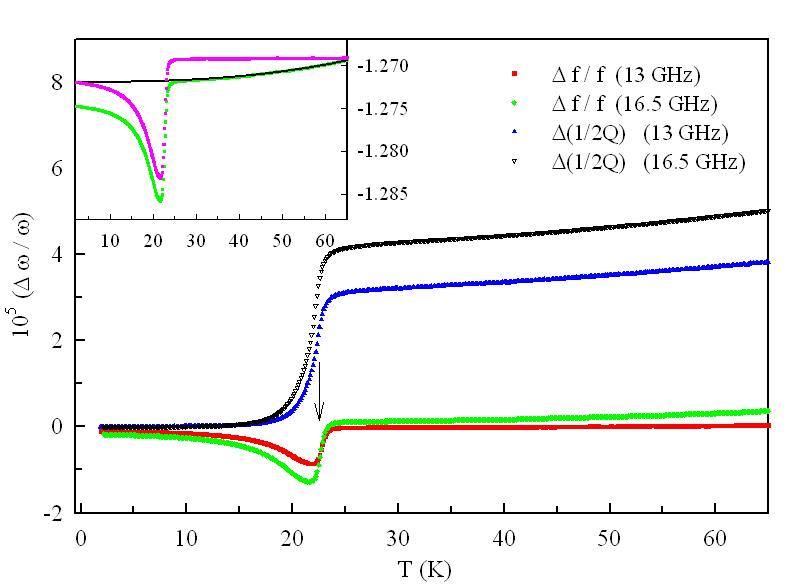} \caption{\label{fig:fig1}
(Color online) Temperature dependence of $\Delta f/f$ and
$\Delta(1/2Q)$ at 13 and 16.5 GHz for sample A. Inset: correction
of the $\Delta f/f$ data for the TE$_{102}$ mode at 16.5GHz.}
\end{figure}

Indeed, a constant $\Gamma$ factor implies identical temperature
dependence of the resonance frequency for an empty cavity and one
loaded with a perfect conductor. The TE$_{101}$ resonance mode at
13 GHz presents only one electric field lobe and the sample which
is located at its center perturbs the mode in a symmetrical way;
we thus expect a temperature dependence of the resonance
frequency very similar to an empty cavity and, thus, a constant
$\Gamma$. The situation is more complex for the TE$_{102}$ mode
since the sample is located in only one of the two lobes yielding
an asymmetrical perturbation of the mode. The resonance frequency
as a function of temperature is thus expected to be modified
relative to the empty cavity and this yields a $\Gamma(T)$ which
explains the non monotonous temperature behavior observed at 16.5
GHz. We should recall that the temperature dependence of the
resonance frequency is due to a progressive increase of the
cavity volume above 20 K.

Since the comparison of the complex penetration depth
$\Tilde{\lambda}$(T) at two frequencies will validate our
approach, it is important to correct the $\Delta f/f$ data at
16.5 GHz . We show this correction in the inset of Fig. 1. Since
the perturbation of the mode is mainly observed above 20 K, we
use a polynomial fit to mimic these effects which are subsequently
substracted from the original data. If the thin film alone
constituted the sample, the constant $\Gamma$ could be calculated
with sufficient precision by associating its geometry to an
elongated ellipsoid. However, the $LaAlO_3$ substrate makes
$\Gamma$ an undetermined parameter depending on the particular
sample and on the resonance mode of the cavity. For each sample,
we thus define two parameters $\Gamma_{13}$ and $\Gamma_{16.5}$
keeping in mind that these parameters are related since the value
of the magnetic penetration depth $\lambda(0) = \lambda_1(0)$
must be frequency independent. We can thus eliminate one degree
of freedom by adjusting accordingly the ratio of the two
parameters, whose values are chosen to yield the best coincidence
of $\lambda_1$ and $\lambda_2$ in the skin depth regime for $T >
T_c$ extracted from the data at 13 and 16.5 GHz ($\lambda_1 =
\lambda_2 = \delta/2$ with the skin depth $\delta =
(2/\omega\sigma_1\mu_0)^{1/2}$). Correction to the $\Delta(1/2Q)$
shifts at 16.5 GHz is not necessary.

\begin{figure}[H,h]
\includegraphics[width=8.5cm]{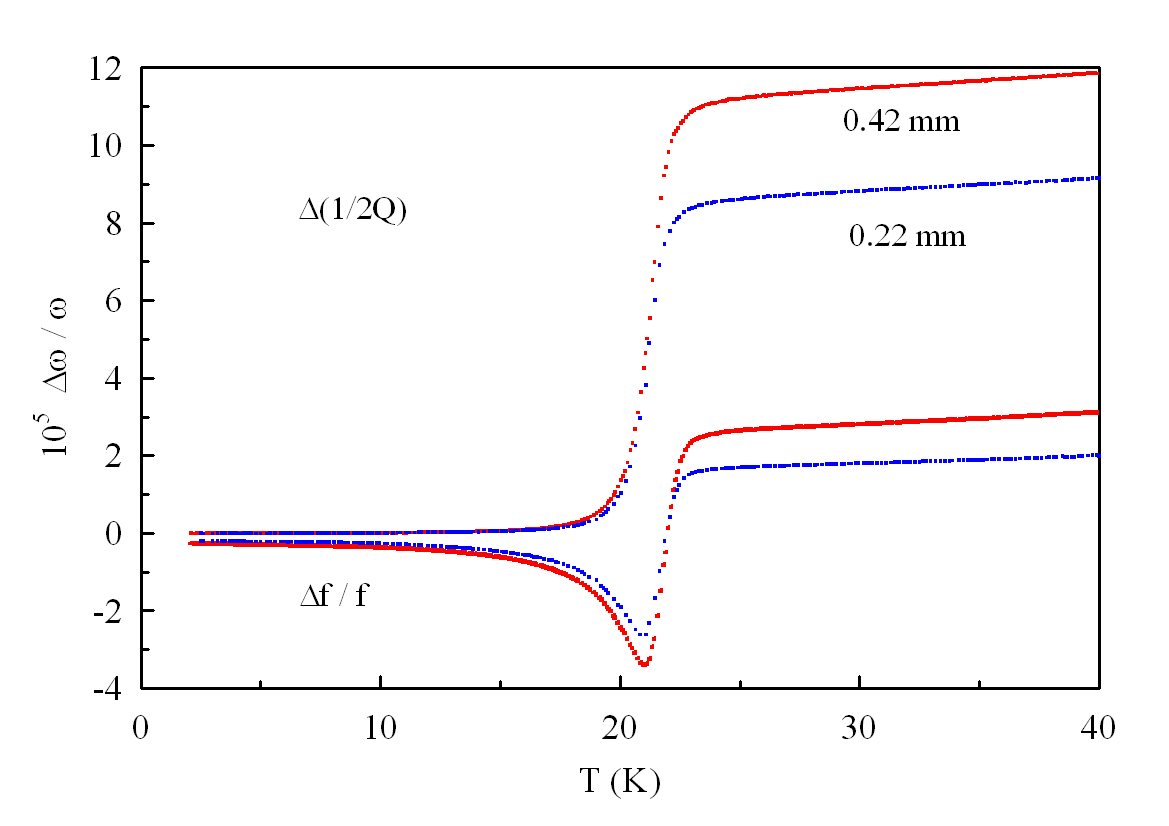} \caption{\label{fig:fig2}
(Color online) Comparison of the complex relative frequency
shifts for two substrate thickness 0.42 mm and 0.22 mm at 16.5 GHz
for the FS configuration.}
\end{figure}

\subsection{Dielectric substrate contribution}

Partly included in the parameter $\Gamma$, the substrate modifies
also the amplitude of the complex frequency shift relative to the
unperturbed state. We propose here a modified procedure adapted
to our PCCO films which is based on the comparison of results
obtained for different substrate configurations: i) reduced
thickness of the substrate; ii) symmetrical sample,
substrate-film-substrate. In Fig. 2, we compare the relative
frequency shifts at 16.5 GHz obtained for thin film B on a 0.42 mm
thick substrate which has been reduced to 0.22 mm after
polishing. The increase of thickness does not affect the
temperature dependence of the shifts; it acts merely as an
amplification factor. Would the situation be similar if the
substrate was present on both sides of the film? A second
substrate with identical dimensions was mechanically added onto
the film. The relative frequency shifts for this configuration
$\it{substrate}$-$\it{film}$-$\it{substrate}$ (SFS) are compared
to the $\it{film}$-$\it{substrate}$ (FS) one in Fig. 3 for the two
resonance modes. The temperature dependence of the shifts is again
maintained but the amplification factor is increased. As
expected, the presence of the substrate increases the frequency
shift and decreases the quality factor as can be easily observed
above $T_c$.

\begin{figure}[H,h]
\includegraphics[width=8.5cm]{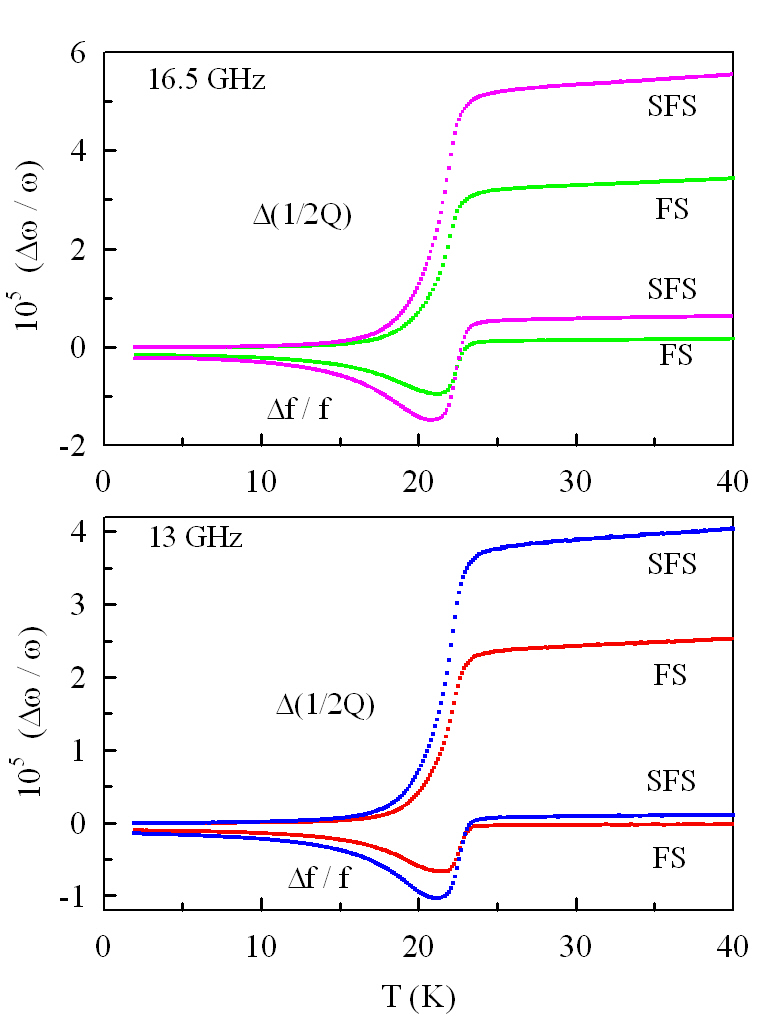} \caption{\label{fig:fig3}
(Color online) Comparison of the complex relative frequency
shifts for the FS and SFS configurations at 13 GHz and 16.5 GHz.}
\end{figure}

The data of Fig. 2 and Fig. 3 suggest to relate the substrate
effects to a scaling factor in the expression connecting the
complex frequency shift to the physical properties of the film.
This scaling factor is found by considering the electric field
configuration of Fig. 4: the substrate on each side of the film
acts as an amplifier of the electric field $\bf{E_s}$ at the
interface, a factor $\zeta$ on the left and $\eta$ on the right.
These factors depend on the thickness and the dielectric constant
of the substrate. If we use these new interface conditions to find
the electric field profile for the slab geometry and the induced
fields $\bf{\tilde D(z)}$ and $\bf{\tilde H(z)}$ (see appendix),
we obtain the following complex frequency shift
\begin{align}
\frac{\Delta\Tilde{\omega}}{\omega} = \kappa \frac{\alpha}{N}
\left[ 1 - \left[ \frac{tanh(\tilde{\gamma}d/2)}{\Tilde{\gamma}
{\omega}^2 \mu_0 \epsilon_0 d/2} + 1 \right]N\right]^{-1}
\end{align}
where $\Tilde{\gamma} = 1 / \Tilde{\lambda}$ is the complex wave
vector. This equation is equivalent to the one obtained for the
complex frequency shift from the perfect conductor to an
arbitrary state (eq.(18) in \cite{PELIGRAD2001}) except for the
scaling factor $\kappa$ which is related to the field
amplification factors by:
\begin{align}
\kappa = {\frac{(\zeta + \eta)}{4}}^2 \left[ 1 -
\frac{1}{2}\left[\frac{\zeta - \eta}{\zeta + \eta}\right]^2\right]
\end{align}
This scaling factor $\kappa$ is evaluated by considering the
experimental data for the FS configuration (Fig. 2) and the SFS
one (Fig. 3) which suggest to write
\begin{align}
\left(\frac{\Delta\Tilde{\omega}}{\omega}\right)_{FS} &=
\kappa_{FS} \left(\frac{\Delta\Tilde{\omega}}{\omega}\right)_{F}
\\ \notag
\\
 \left(\frac{\Delta\Tilde{\omega}}{\omega}\right)_{SFS}
&= \kappa_{SFS}
\left(\frac{\Delta\Tilde{\omega}}{\omega}\right)_{F}
\end{align}
where we have defined the relative complex frequency shift of the
film (F) alone when $\kappa$ = 1. If we define $\beta$ as the
ratio $\kappa_{FS}/\kappa_{SFS}$ and we put $\zeta = \eta = 1 +
\epsilon$ valid for the data of Fig. 3 (same substrate geometry
on both sides of the film), we write the scaling factors as
\begin{align}
\kappa_{FS} = {\frac{(2 + \epsilon)}{4}}^2 \left[ 1 -
\frac{1}{2}\left[\frac{\epsilon}{2 + \epsilon}\right]^2\right],
\kappa_{SFS} = (1 + \epsilon)^2
\end{align}
This last result allows us to express the factor $\epsilon$,
which represents the deviation from the $\it{no}$-$\it{substrate}$
configuration, as a function of $\beta =\kappa_{FS}/\kappa_{SFS}$
\begin{align}
\epsilon = \frac{\beta - 2 + \sqrt{{(\beta -2)^2 - (2 - \beta /
4)(1 - \beta)}}}{2 - \beta / 4}
\end{align}
Because the undetermined constant $\Gamma$, defining $\Delta f/f$,
is an adjustable parameter in the final data treatment, we use the
$\Delta(1/2Q)$ shifts of Fig. 3 to determine the factor $\beta$:
practically the same value, 1.59 and 1.62, is respectively
obtained at 13 GHz and 16.5 GHz which yields $\epsilon \approx
0.65$. There is, of course, a dependence of the parameter
$\epsilon$ on the thickness of the substrate as suggested in Fig.
2. The best fit is obtained with a value of $\epsilon$ reduced by
a factor 0.43 when the thickness has been decreased by a factor
0.53. This last result suggests a quasi-linear relationship
between $\epsilon$ and the thickness of the substrate at these
frequencies.

\begin{figure}[H,h]
\includegraphics[width=8.5cm]{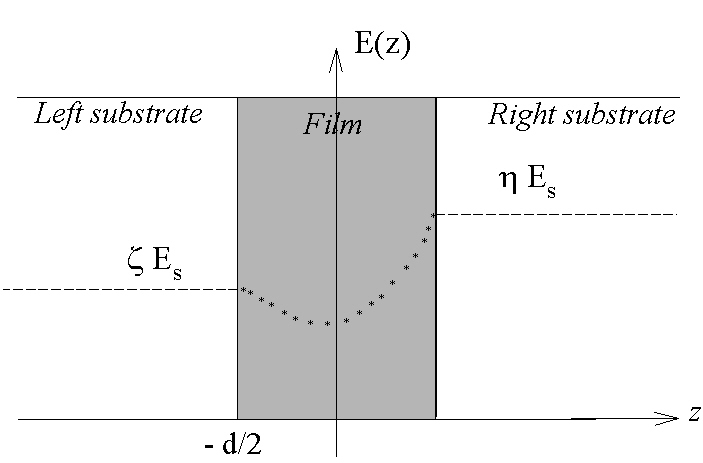} \caption{\label{fig:fig4}
(Color online) Electric field profile as a function of the
position z from the center of the film of thickness \textit{d}.
E$_s$ is the electric field outside the substrate and $\zeta$ and
$\eta$ the amplification factors of the substrate on both sides
of the films.}
\end{figure}

\subsection{Complex penetration depth : d-wave symmetry}

In the preceding sections we have described the tools necessary
to extract the complex penetration depth $\Tilde{\lambda}$ and
subsequently the conductivity $\Tilde{\sigma}$ from our relative
frequency shifts data. By adjusting the constant $\Gamma$ and
using the factor $\kappa$ determined from Eq.(5) for the FS
configuration , we search for each frequency the best
superposition of $\lambda_1$ and $\lambda_2$ in the normal state
extracted from Eq.(1).

\begin{figure}[H,h]
\includegraphics[width=8.5cm]{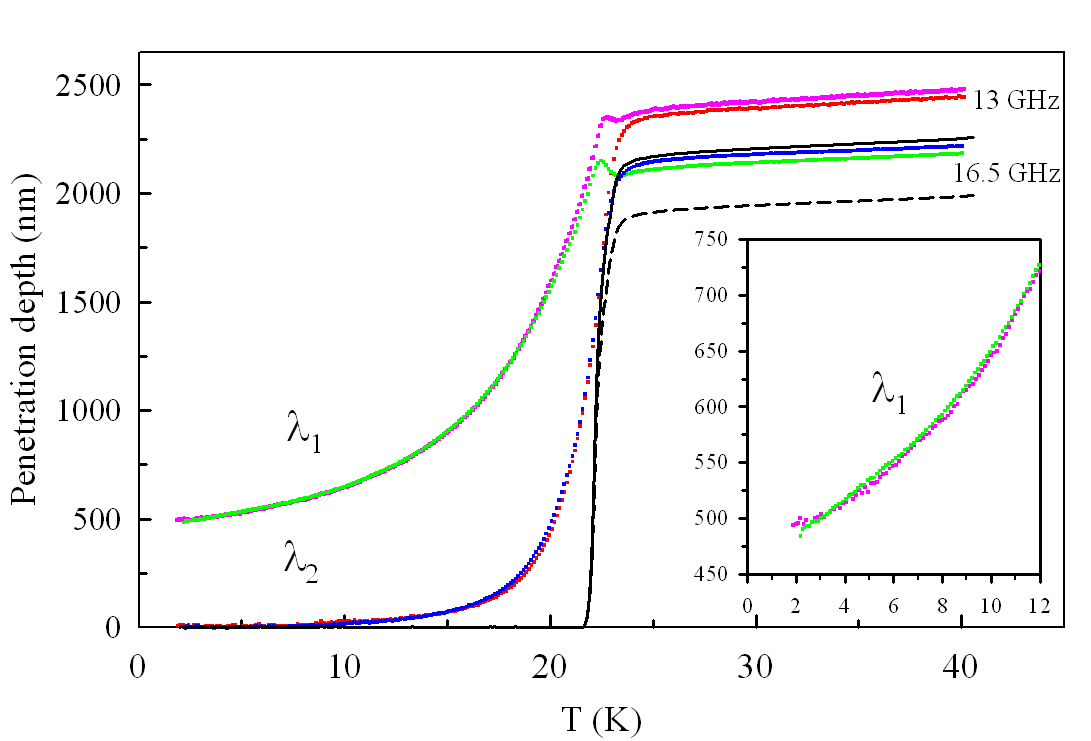} \caption{\label{fig:fig5}
(Color online) Complex penetration depth of thin film A as a
function of temperature at 13 and 16.5 GHz. Continuous and broken
lines represent the $\delta/2$ calculated from DC resistivity
data.}
\end{figure}

We show in Fig. 5 the complex penetration depth of the thin film A
between 2 K and 40 K at 13 GHz and 16.5 GHz. This measurement is
analogous  to a surface impedance one that could have been
performed on a single crystal. However, thin films offer the
advantage of obtaining directly the absolute value $\lambda(0)$.
In the normal state for $T > T_c$, the superposition of
$\lambda_1$ and $\lambda_2$ is very good for both frequencies,
although better at 13 GHz since no correction of the $\Delta f/f$
data is necessary. As expected the penetration depth is larger at
13 than at 16.5 GHz in the normal state, in agreement with the
frequency dependence of the skin depth $\delta$. The monotonous
increase of the real and imaginary parts with temperature
compares very well with the $\delta/2$ plots obtained with the DC
resistivity measured on the same film and also presented in Fig.
5. The microwave value is just a little larger than the DC one
and this is explained by uncertainties on the parameters
$\alpha$, $N$ and $\kappa$ appearing in Eq.(1). In the
superconducting state, we observe no frequency effect on
$\lambda_1$ on a wide temperature range although we imposed the
coincidence at the lowest temperatures only. The extrapolated
value towards zero temperature is $\lambda (0) = 460 \pm 100$
nanometers; the large uncertainty is attributed to the adjustment
of the scaling factor $\kappa$. This value is larger than the one
found in the literature \cite{LEMBERGER}, but is consistent with
a higher normal state resistivity. When the temperature is
increased from 2 K, $\lambda_1$ increases quasi-linearly first up
to 6-7 K ($\sim$ 17 nm/K) and quadratically after (inset of Fig.
5); then, it goes through a small maximum just below $T_c$. Such a
signature is also observed in other conventional \cite{POIRIER}
and unconventional superconductors \cite{ORMENO}. When $T_c$ is
approached from above, the initial reduction of quasiparticle
density leads to a first order reduction in screening from
quasiparticles ($\delta$) but only a second order increase in
screening from the superfluid fraction. This leads to a peak that
moves to lower temperatures on increasing frequency as observed
in Fig. 5. In the superconducting state, the imaginary part
$\lambda_2$ expresses a rapid reduction of screening by
quasiparticles below $T_c$.

\begin{figure}[H,h]
\includegraphics[width=8.5cm]{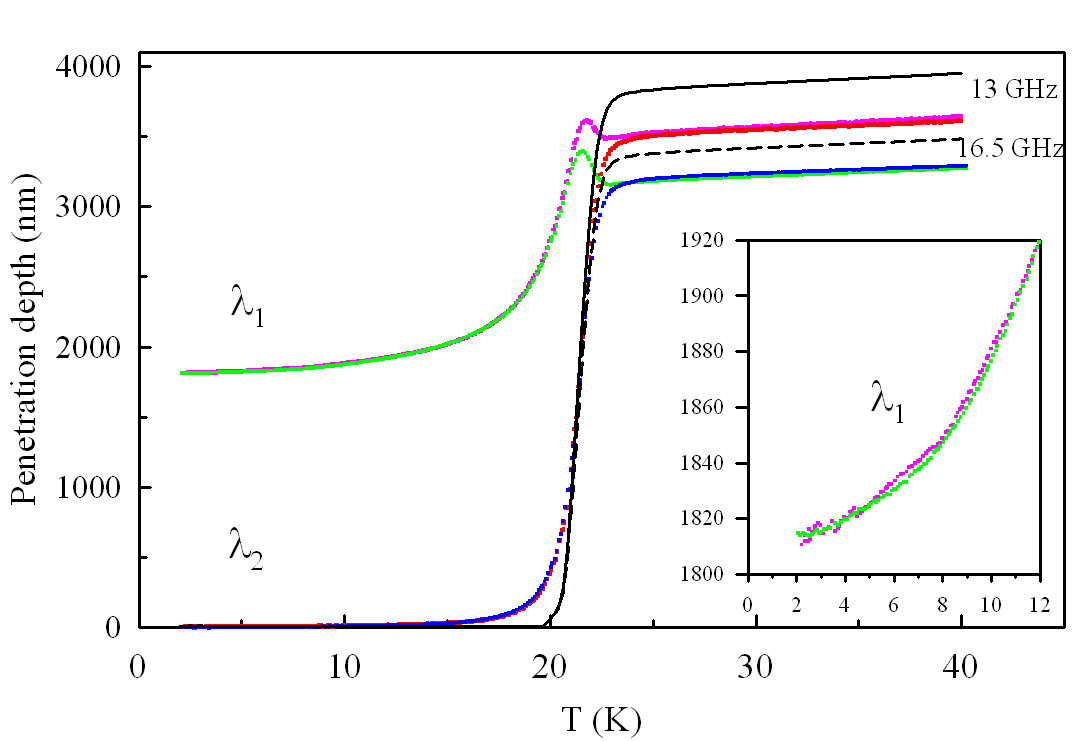} \caption{\label{fig:fig6}
(Color online) Complex penetration depth of thin film B as a
function of temperature at 13 and 16.5 GHz. Continuous and broken
lines represent the $\delta/2$ calculated from DC resistivity
data.}
\end{figure}

The complex penetration depth of the thin film B between 2 K and
40 K at 13 GHz and 16.5 GHz is presented in Fig. 6. For this
sample, the superposition of $\lambda_1$ and $\lambda_2$ in the
normal state is excellent at the two frequencies. Here the
absolute values are a little smaller than the $\delta/2$ data
calculated from the DC resistivity. In the superconducting state,
no frequency effects are observed on a wide temperature range on
the real part $\lambda_1$ with an extrapolated value $\lambda(0) =
1850 \pm 150$ nanometers. When the temperature is increased from
2 K, $\lambda_1$ increases quadratically (inset of Fig. 6, $\sim$
6 nm/K) and a small peak is still observed just below $T_c$. The
imaginary part $\lambda_2$ appears to decrease more rapidly below
$T_c$ for this film compared to the previous one, the reduction
of quasiparticle screening being more efficient. This is
consistent with the data in Table 1 showing that film A had a
higher quality than film B. This is confirmed by a much lower
penetration depth $\lambda (0)$ and a quasi-linear increasing
rate at low temperatures compared to a quadratic one when
impurity scattering is more important. Thus, these results appear
to be consistent with a $\it{d}$-wave order parameter with lines
of nodes as already suggested by several other studies.

\begin{figure}[H,h]
\includegraphics[width=8.5cm]{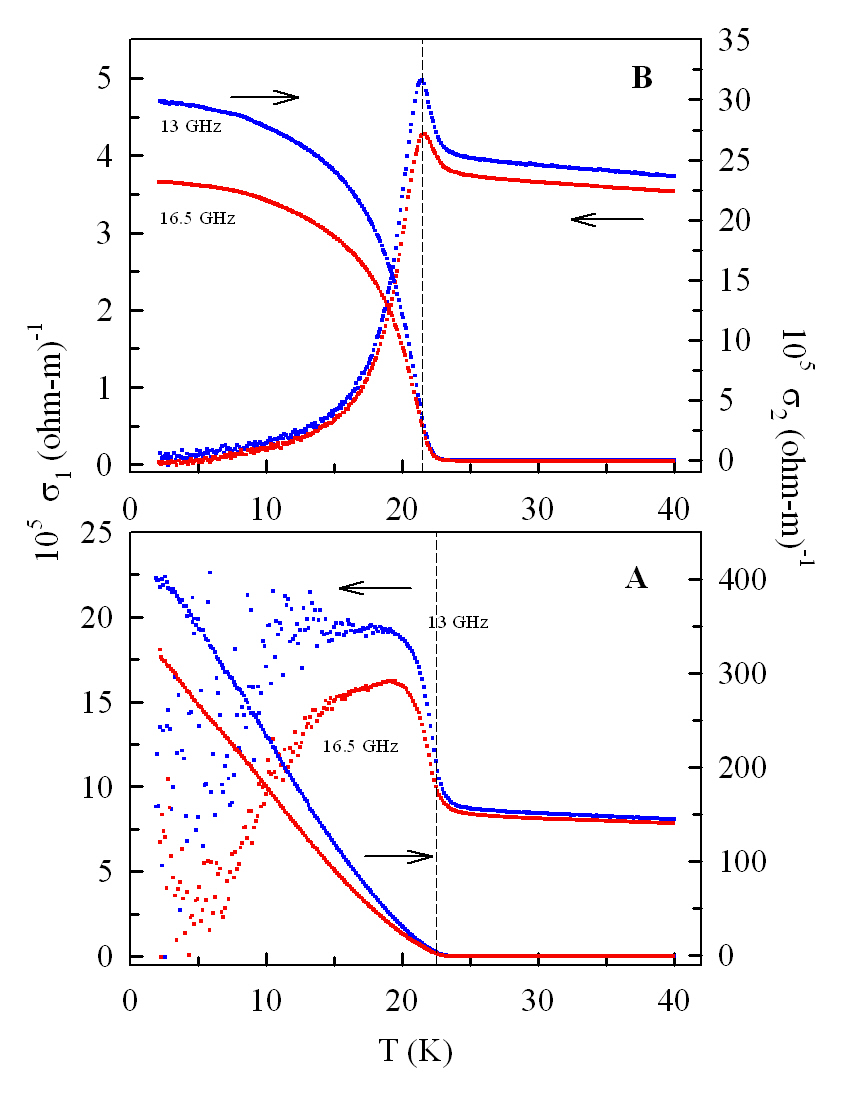} \caption{\label{fig:fig7}
(Color online) Complex microwave conductivity of thin film A and B
as a function of temperature at 13 GHZ (blue) and 16.5 GHz (red).
The superconducting critical temperature is indicated by a dashed
line.}
\end{figure}

\subsection{Complex conductivity : two-fluid behavior}

Our microwave penetration depth data are most revealing when
transformed into complex conductivity which is presented for both
films in Fig. 7. According to the two-fluid model, the real part
$\sigma_1 (T)$ is related to quasiparticle effects and the
imaginary part $\sigma_2 (T)$ to the superfluid density. In the
normal state, the latter is zero when the former is related to a
constant quasiparticle density $n_n$ and a temperature dependent
relaxation time $\tau (T)$. We observe these features in Fig. 7
for both thin films; the temperature dependence of $\sigma_1 (T)$
is only due to the relaxation time $\tau (T)$ which increases
smoothly with decreasing temperature. The small difference
observed between the 13 GHz and 16.5 GHz data do not represent
frequency effects; they are due to an imperfect coincidence of
$\lambda_1$ and $\lambda_2$ (Fig. 5 and Fig. 6) when adjusting
the parameter $\Gamma$. In the superconducting state, the
behavior of the conductivity is highly sample dependent and we
discuss these data separately.

For film A with the best quality, $\sigma_1$ increases first
sharply around $T_c$ when the superconducting temperature is
approached from above, then presents a wide maximum before
decreasing further toward zero at the lowest temperatures. An
increase in frequency appears to suppress slightly the maximum.
The imaginary part $\sigma_2$ begins to increase weakly above
$T_c$ with a faster rate at the superconducting temperature and
it goes through a quasi-linear regime below 15 K at both
frequencies. This dependence is consistent with a linear increase
of the superfluid density $n_s(T)$ at low temperatures, $\sigma_2
(T) = n_s(T)e^2/m^*\omega$ ($m^*$ being the effective mass). Such
a linear behavior was never clearly observed on ${\lambda
(T)}^{-2}\sim n_s$ for optimally doped PCCO thin films
\cite{LEMBERGER}.

Enhanced conductivity effects around $T_c$ are related to
superconducting fluctuations which yield generally a peak at
$T_c$ on $\sigma_1$ (paraconductivity); however such a peak
merges with the wide maximum appearing below $T_c$ which is
likely due to a decrease of the relaxation rate of the
quasiparticles. These two features appear on different
temperature scales in YBCO allowing them to be separated
\cite{BONN}. In our case for the PCCO thin films, their proximity
makes them impossible to separate. Moreover, the gradual
suppression of the wide maximum with increasing frequency is
fully consistent with the relaxation rate scenario discussed
above.

The superconducting state conductivity of thin film B is quite
different even if the superconducting temperature is lower by
only 1 K. $\sigma_1 (T)$ presents a small peak confined around
$T_c$ and then it decreases rapidly in the superconducting state
without showing a frequency dependence. At the same time,
$\sigma_2 (T)$ increases steeply with only a small smearing
around $T_c$. Below 15 K, both parts of the conductivity vary
quadratically with temperature. The small peak on $\sigma_1$ and
the weak smearing of $\sigma_2$ at $T_c$ are due to much weaker
superconducting fluctuations for this film. The absence of a wide
maximum and frequency effects on $\sigma_1$ below $T_c$, and a
similar temperature dependence of the real and imaginary parts
below 15 K indicate that only density effects intervene in the
quasiparticle contribution. Their relaxation rate which is
dominated by impurity scattering does not appear modified
substantially in the superconducting state. Impurity scattering
also modifies the linear regime of the superfluid density
observed in film A to a quadratic one in film B.

These microwave conductivity results appear consistent with a
$\it{d}$-wave order parameter in these electron-doped cuprates.
On the one hand, the highest quality film (A) can be treated in
the $\it{clean}$ limit according to the quasi-linear behavior of
the penetration depth and of the superfluid density at low
temperatures. Concomitantly, quasiparticle effects due to a
modified relaxation rate are observed. On the other hand, a film
of a lesser quality (B) goes rapidly in the $\it{dirty}$ limit
yielding a quadratic temperature dependence for $\lambda (T)$ and
$n_s(T)$ over the same temperature range. Moreover, quasiparticle
relaxation rate effects have practically disappeared when
superconducting fluctuations, although reduced, still enhance the
conductivity at $T_c$. Compared to film A, the superfluid density
appears to be reduced by an order of magnitude in film B although
they were grown with the same ablation procedure. Film B, which is
substantially thinner than film A, should be more affected by the
quality of the interface with the substrate, interface which is
likely the source of the increased impurity scattering rate.

\section{CONCLUSION}
In this paper, we have shown that a microwave cavity perturbation
technique can be used to deduce appropriately the absolute value
of the penetration depth and its related complex conductivity as
a function of temperature for thin films of electron-doped
cuprates in their superconducting state. We have treated the thin
films, located in the maximum electric field for two resonance
modes, using the general solution for the complex frequency shift
of Peligrad et $\textit{al.}$ We have shown how to characterize
the $\textit{unperturbed state}$, which is essential to the
determination of the appropriate frequency shifts, and how to
take into account the dielectric substrate effects in the data
treatment. Then, with only one adjustable parameter $\Gamma$ and
a pre-determined scaling factor $\kappa$, we have deduced the
complex $\tilde{\lambda} (T)$ and $\tilde{\sigma}(T)$ for two
Pr$_{1.85}$Ce$_{0.15}$CuO$_{4-\delta}$ thin films of slightly
different quality. These microwave results reproduce the main
characteristics of the superconducting state found in the
hole-doped cuprates. We observe a quasi-linear behavior of
$\lambda (T)$ at low temperatures with a crossover to a quadratic
dependence when impurity scattering becomes more important. An
important decrease of the quasiparticle scattering rate below
$T_c$ must be invoked to interpret the wide conductivity maximum
below $T_c$; this quasiparticle effect is rapidly quenched by
impurity scattering. Finally, important superconducting
fluctuations are found around $T_c$ and they could be at the
origin of the peculiar temperature dependence of the superfluid
density observed in the high quality sample.

\acknowledgments{The authors thank M. Castonguay and S. Pelletier
for their technical support. This work was supported by grants
from the Fonds Qu\'eb\'ecois de la Recherche sur la Nature et les
Technologies (FQRNT) and from the Natural Science and Engineering
Research Council of Canada (NSERC).}

\section*{APPENDIX: CONTRIBUTION FROM THE SUBSTRATE}

To find the spatial dependence of the electric field inside a
film having a dielectric substrate on both sides, we start from
the following dependence for the slab geometry with the complex
wave vector $\tilde{\gamma} = (1 + i)/\delta = 1/\tilde{\lambda}$.
\begin{align}
\Tilde{E}(z) &= \Tilde{A} e^{-\tilde{\gamma}z} + \Tilde{B}
e^{+\tilde{\gamma}z}  \tag{A.1}
\end{align}
We then apply the boundary conditions at both interfaces as
illustrated in Fig. 4
\begin{align}
\Tilde{E}(d/2)& = \eta \tilde{E}_s = \Tilde{A}
e^{-\tilde{\gamma}d/2} + \Tilde{B} e^{+\tilde{\gamma}d/2}\notag \\
\Tilde{E}(-d/2)& = \zeta \tilde{E}_s = \Tilde{A}
e^{+\tilde{\gamma}d/2} + \Tilde{B} e^{-\tilde{\gamma}d/2}\tag{A.2}
\end{align}
and we obtain the following spatial profiles for the electric
$\tilde{E}(z)$ and magnetic $\tilde{B}(z)$ fields.
\begin{align}
\Tilde{E}(z)= \frac{\tilde{E}_s}{\sinh{(\tilde{\gamma}d})}[&(\eta
e^{+\tilde{\gamma}d/2} - \zeta
e^{-\tilde{\gamma}d/2})e^{+\tilde{\gamma}z} \notag
 \\
&+ (\zeta e^{+\tilde{\gamma}d/2} - \eta
e^{-\tilde{\gamma}d/2})e^{-\tilde{\gamma}z}] \notag
\end{align}
\begin{align}
\Tilde{B}(z)=
\frac{\tilde{\gamma}\tilde{E}_s}{2i\omega\sinh{(\tilde{\gamma}d})}[&(\eta
e^{+\tilde{\gamma}d/2} - \zeta
e^{-\tilde{\gamma}d/2})e^{+\tilde{\gamma}z} \notag
 \\
&- (\zeta e^{+\tilde{\gamma}d/2} - \eta
e^{-\tilde{\gamma}d/2})e^{-\tilde{\gamma}z}] \tag{A.3}
\end{align}
Then, we calculate the induced fields $\tilde{D}(z)$ and
$\tilde{H}(z)$ inside the slab.  Since $\tilde{E}(z)$ is
different at the two interfaces, we may assume that the field
$\tilde{D}(z)$ could vary linearly with the distance $z$
according to the following relation
\begin{align}
\tilde{D}(z) = {\tilde{\epsilon}}_s \epsilon_0
\tilde{E}_s\left[\frac{(\eta-\zeta)}{d}z +
\frac{(\eta+\zeta)}{2}\right]\tag{A.4}
\end{align}
This will yield a magnetic field varying as
\begin{align}
\tilde{H}(z) = i \omega {\tilde{\epsilon}}_s \epsilon_0
\tilde{E}_s\left[\frac{(\eta-\zeta)}{2d}z^2 +
\frac{(\eta+\zeta)}{2}z \right]\tag{A.5}
\end{align}
with an amplitude determined at the interfaces with the condition
\begin{align}
\tilde{H}(d/2) - \tilde{H}(-d/2) = i \frac{(\eta+\zeta)}{2}\omega
{\tilde{\epsilon}}_s \epsilon_0 \tilde{E}_sd \tag{A.6}
\end{align}
With this approximation, we find that the relative permittivity
of the sample is not modified by the substrate and we still have
the relation.
\begin{align}
\tilde{\epsilon}_s =
\frac{-2\tilde{\gamma}}{\omega^2\tilde{\mu}_r\mu_0\epsilon_0d}\tanh{(\tilde{\gamma}d/2)}
\tag{A.7}
\end{align}
We can now use these expressions for the electric and magnetic
fields inside the sample, $\tilde{E}(z)$, $\tilde{B}(z)$,
$\tilde{D}(z)$ and $\tilde{H}(z)$ to calculate the complex
frequency shift $\Delta\tilde{\omega}/\omega$ relative to the
perfect conductor state (unperturbed state) with the equation (5)
of ref.\cite{PELIGRAD2001} to obtain
\begin{align}
\frac{\Delta\tilde{\omega}}{\omega} = - {\frac{(\zeta +
\eta)}{4}}^2  \left[ 1 - \frac{1}{2}\left[\frac{\zeta -
\eta}{\zeta + \eta}\right]^2\right]\frac{\alpha}{N} \notag \\
\left[- \frac{1}{1+(\tilde{\epsilon}_s - 1)N}\right] \tag{A.8}
\end{align}
Since the relative permittivity $\tilde{\epsilon}_s$ (A.7) is not
affected by the substrate within our approximation, we finally
obtain the following expression for the relative frequency shift
\begin{align}
\frac{\Delta\Tilde{\omega}}{\omega} = \kappa \frac{\alpha}{N}
\left[ 1 - \left[ \frac{tanh(d/2\Tilde{\lambda})}{\Tilde{\gamma}
{\omega}^2 \mu_0 \epsilon_0 d/2} + 1 \right]N\right]^{-1}
\tag{A.9}
\end{align}
\begin{align}
\kappa = {\frac{(\zeta + \eta)}{4}}^2 \left[ 1 -
\frac{1}{2}\left[\frac{\zeta - \eta}{\zeta +
\eta}\right]^2\right]\tag{A.10}
\end{align}
where the substrate appears only as a scaling factor $\kappa$.
This equation for the frequency shift is identical to Eq.(18) of
Ref.\cite{PELIGRAD2001}, except for the scaling factor $\kappa$.

When the film thickness is smaller than the penetration depth
$\tilde{\lambda}$, we can approximate
$\tan(d/(2\tilde\lambda)\approx d/2\tilde\lambda$. The complex
penetration depth is finally obtained from the relative frequency
shift.
\begin{align}
\tilde{\lambda} = \left[\frac{\omega^2\epsilon_0\mu_0}{N}\left[1 -
\frac{\alpha}{N\Delta\tilde{\omega}/\omega}\kappa-N\right]\right]^{-1/2}\tag{A.11}
\end{align}


\begin{thebibliography}{23}

\bibitem{CASIMIR} H.B.G. Casimir, Philips Res. Rep. \textbf{6},
162 (1951).
\bibitem{WALDRON} R.A. Waldron, \textit{Theory of Guided
Electromagnetic Waves} (Van Nostrand, Reinhold Co., London, 1969).
\bibitem{HALBRITTER} J. Halbritter, J. Appl. Phys. \textbf{41},
4581 (1970).
\bibitem{BONN1993} D.A. Bonn, Ruixing Liang, T.M. Riseman, D.J.
Baar, D.C. Morgan, Kuan Zhang, P. Dosanjh, T.L. Duty, A.
MacFarlane, G.D. Morris, J.H. Brewer and W.N. Hardy, Phys. Rev. B
\textbf{47}, 11314 (1993).
\bibitem{HARDY1993} W.N. Hardy, D.A. Bonn, D.C. Morgan, R. Liang
and K. Zhang, Phys. Rev. Lett. \textbf{70}, 3999 (1993).
\bibitem{JACOBS1995} T. Jacobs, S. Sridhar, Q. Li, G.D. Gu and N.
Koshizuka, Phys. Rev. Lett. \textbf{75}, 4516 (1995).
\bibitem{LEE1996} S.-F. Lee, D.C. Morgan, R.J. Ormeno, D.M. Broun,
R.A. Doyle, J.R. Waldram and K. Kadowaki, Phys. Rev. Lett.
\textbf{77}, 735 (1996).
\bibitem{KITANO1998} H. Kitano, T. Hnaguri and A. Maeda, Phys.
Rev. B \textbf{57}, 10496 (1998).
\bibitem{KOKALES} J.D. Kokales, P. Fournier, L.V. Mercaldo, V.V.
Talanov, R.L. Greene and S.N. Anlage, Phys. Rev. Lett.
\textbf{85}, 3696 (2000).
\bibitem{PROZOROV} R. Prozorov, R.W. Gianetta, P. Fournier and
R.L. Greene, Phys. Rev. Lett. \textbf{85}, 3700 (2000).
\bibitem{SNEZHKO} A. Snezhko, R. Prozorov, D.D. Lawrie, R.W.
Gianetta, J. Gauthier, J. Renaud and P. Fournier, Phys. Rev.
Lett. \textbf{92}, 157005 (2004).
\bibitem{SCHAUMBURG1992}  G. Schaumburg, H.W. Helberg, P.
Berberich and H. Kinder, Ann. Physik \textbf{1}, 584 (1992).
\bibitem{SCHAUMBURG1994} G. Schaumburg and H.W. Helberg, J. Phys
III France \textbf{4}, 917 (1994).
\bibitem{PELIGRAD1998} D.-N. Peligrad, B. Nebendahl, C. Kessler,
M. Mehring, A. Dulcic, M. Pozek and D. Paar, Phys. Rev. B
\textbf{58}, 11652 (1998).
\bibitem{PELIGRAD2001} D.-N. Peligrad, B. Nebendahl, M. Mehring,
A. Dulcic, M. Pozek and D. Paar, Phys. Rev. B \textbf{64}, 224504
(2001).
\bibitem{SHCHEGOLEV} L.I. Buravov and I.F. Shchegolev, Instrum.
Exp. Tech. \textbf{14}, 528 (1971).
\bibitem{KLEIN} O. Klein, S. Donovan, M. Dressel and G. Gruner,
Int. J. Infrared Millim. Waves, \textbf{14}, 2423 (1993).
\bibitem{FOURNIER1998} P. Fournier, P. Mohanty, E. Maiser, S.
Darzens, T. Venkatesan, C.J. Lobb, G. Czjzek, R.A. Webb and R.L.
Greene, Phys. Rev. Lett. \textbf{81}, 4720 (1998).
\bibitem{LEMBERGER}M.S. Kim, J.A. Skinta, T.R. Lemberger, A.
Tsukada and M. Naito, Phys. Rev. Lett. \textbf{91}, 087001
(2003); J.A. Skinta and T.R. Lemberger, Phys. Rev. Lett.
\textbf{88}, 207003 (2002).
\bibitem{POIRIER} A similar peak is observed on the imaginary part
of the 16.5 GHz surface impedance of superconducting zinc ($T_c$
= 0.85 K).
\bibitem{ORMENO} R. J. Ormeno, A. Sibley and C. E. Gough, Phys.
Rev. Lett. \textbf{88}, 047005 (2002).
\bibitem{BONN} A. Hosseini, R. Harris, S. Kamal, P. Dosanjh, J.
Preston, R. Liang, W.N. Hardy and D.A. Bonn, Phys. Rev.
\textbf{B}, 1349 (1999).


\end{thebibliography}
\end{document}